\documentclass[sigconf]{acmart}
\AtBeginDocument{%
  \providecommand\BibTeX{{%
    \normalfont B\kern-0.5em{\scshape i\kern-0.25em b}\kern-0.8em\TeX}}}

\usepackage{amsmath}
\usepackage{amsthm}
\usepackage{amsfonts}
\usepackage{bbm}
\usepackage{bm}
\usepackage{xcolor}
\usepackage{pgf}
\usepackage{import}
\usepackage{multirow,booktabs,makecell}
\usepackage{tikz}
\usetikzlibrary{positioning}
\usepackage{algorithmic}
\usepackage{textcomp}
\usepackage{caption}
\usepackage[linesnumbered,ruled,noend]{algorithm2e}

\SetCommentSty{mycommfont}

\usepackage{float}
\usepackage{bm}

\definecolor{c0}{HTML}{CC0033}
\definecolor{c1}{HTML}{991A59}
\definecolor{c2}{HTML}{663380}
\definecolor{c3}{HTML}{334DA6}
\definecolor{c4}{HTML}{0066CC}

\newtheorem{theorem}{Theorem}

\newtheorem{innercustomthm}{Theorem}
\newenvironment{customthm}[1]
  {\renewcommand\theinnercustomthm{#1}\innercustomthm}
  {\endinnercustomthm}

\let\originalleft\left
\let\originalright\right
\renewcommand{\left}{\mathopen{}\mathclose\bgroup\originalleft}
\renewcommand{\right}{\aftergroup\egroup\originalright}

\newcommand*\makealph[1]{\symbol{\numexpr96+#1}}

\newcommand{\set}[1]{#1}
\newcommand{\neighbors}[1]{\Gamma\left(#1\right)}
\newcommand{\setsize}[1]{\left|#1\right|}
\newcommand{\prob}[1]{\Pr\left(#1\right)}
\newcommand{\jacc}[1]{\mathrm{J}\left(#1\right)}
\newcommand{\adad}[1]{\mathrm{A}\left(#1\right)}
\newcommand{\idf}[1]{\mathrm{idf}\left(#1\right)}
\newcommand{\expected}[1]{\mathbb{E}\left[#1\right]}
\newcommand{\variance}[1]{\mathrm{Var}\left(#1\right)}
\newcommand{\covariance}[1]{\mathrm{Cov}\left(#1\right)}
\newcommand{\indicator}[1]{\bm{1}\left(#1\right)}

\newcommand{\normalcdf}[1]{\Phi\left(#1\right)}
\newcommand{\normalppf}[1]{\Phi^{-1}\left(#1\right)}
\newcommand{\hashint}[1]{h\left(#1\right)}
\newcommand{\hashvec}[1]{\phi\left(#1\right)}
\newcommand{\hashqvec}[1]{\psi\left(#1\right)}
\newcommand{\bigo}[1]{O\left(#1\right)}
\newcommand{\abs}[1]{\left|#1\right|}

\newcommand{\ourhash}{DotHash}

\copyrightyear{2023}
\acmYear{2023}
\setcopyright{rightsretained}
\acmConference[KDD '23]{Proceedings of the 29th ACM SIGKDD Conference on Knowledge Discovery and Data Mining}{August 6--10, 2023}{Long Beach, CA, USA}
\acmBooktitle{Proceedings of the 29th ACM SIGKDD Conference on Knowledge Discovery and Data Mining (KDD '23), August 6--10, 2023, Long Beach, CA, USA}
\acmDOI{10.1145/3580305.3599314}
\acmISBN{979-8-4007-0103-0/23/08}
\begin{document}


\title{DotHash:~Estimating~Set~Similarity~Metrics for~Link~Prediction~and~Document~Deduplication}




\author{Igor Nunes}
\affiliation{%
  \institution{University of California, Irvine}
  \city{Irvine, CA}
  \country{USA}}
\email{igord@uci.edu}

\author{Mike Heddes}
\affiliation{%
  \institution{University of California, Irvine}
  \city{Irvine, CA}
  \country{USA}}
\email{mheddes@uci.edu}

\author{Pere Vergés}
\affiliation{%
  \institution{University of California, Irvine}
  \city{Irvine, CA}
  \country{USA}}
\email{pvergesb@uci.edu}

\author{Danny Abraham}
\affiliation{%
  \institution{University of California, Irvine}
  \city{Irvine, CA}
  \country{USA}}
\email{dannya1@uci.edu}

\author{Alex Veidenbaum}
\affiliation{%
  \institution{University of California, Irvine}
  \city{Irvine, CA}
  \country{USA}}
\email{alexv@ics.uci.edu}

\author{Alex Nicolau}
\affiliation{%
  \institution{University of California, Irvine}
  \city{Irvine, CA}
  \country{USA}}
\email{nicolau@ics.uci.edu}

\author{Tony Givargis}
\affiliation{%
  \institution{University of California, Irvine}
  \city{Irvine, CA}
  \country{USA}}
\email{givargis@uci.edu}







\renewcommand{\shortauthors}{Nunes and Heddes et al.}

\begin{abstract}

Metrics for set similarity are a core aspect of several data mining tasks. To remove duplicate results in a Web search, for example, a common approach looks at the \textit{Jaccard index} between all pairs of pages. In social network analysis, a much-celebrated metric is the \textit{Adamic-Adar index}, widely used to compare node neighborhood sets in the important problem of predicting links. However, with the increasing amount of data to be processed, calculating the exact similarity between all pairs can be intractable. The challenge of working at this scale has motivated research into efficient estimators for set similarity metrics. The two most popular estimators, \textit{MinHash} and \textit{SimHash}, are indeed used in applications such as document deduplication and recommender systems where large volumes of data need to be processed. Given the importance of these tasks, the demand for advancing estimators is evident. We propose \ourhash{}, an unbiased estimator for the intersection size of two sets. \ourhash{} can be used to estimate the Jaccard index and, to the best of our knowledge, is the first method that can also estimate the Adamic-Adar index and a family of related metrics. We formally define this family of metrics, provide theoretical bounds on the probability of estimate errors, and analyze its empirical performance. 
Our experimental results indicate that \ourhash{} is more accurate than the other estimators in link prediction and detecting duplicate documents with the same complexity and similar comparison time.
\end{abstract}




\maketitle

\section{Introduction}
\label{sec:introduction}

Many current challenges---and opportunities---in computer science stem from the sheer scale of the data to be processed~\cite{fan2014challenges, hariri2019uncertainty}. Among these challenges, one of the most outstanding is comparing collections of objects, or simply \textit{sets}\footnote{We shall refer to \textit{sets}, indistinctly, as collections with or without repeated elements.}. This demand arises, for example, in important problems such as comparing text documents or social media profiles. The challenge is often not the size of each set, but the number of pairwise comparisons over a large dataset of sets. This has motivated research on \textit{estimating} existing set similarity measures, the main subject of this paper.

The search for methods to compare sets of elements is long-standing: more than a century ago, \citet{gilbert1884finley} and \citet{jaccard1912distribution} independently proposed a measure that is still widely used, known as the \textit{Jaccard index}. The metric is defined as the ratio between the sizes of the intersection and the union of two sets. 
With the explosion of available data, brought about mainly by the advent of the Web, the Jaccard index has become prevalent as an essential tool in data mining and machine learning. Important applications include information retrieval~\cite{niwattanakul2013using,sefid2019cleaning}, natural language processing~\cite{yih2007improving,strube2006wikirelate}, and image processing~\cite{liu2020deep, padilla2020survey}, among several others~\cite{choi2010survey}.

Another academic field
in which set similarity has become crucial is \textit{network science}. This field studies network representations of physical, biological and social phenomena and is used to understand complex systems in various disciplines~\cite{barabasi2013network}. Such networks are modeled using graphs, a mathematical abstraction tool where sets are ubiquitous. Marked by its growing relevance, this area also gave rise to one of the most famous set similarity metrics: the \textit{Adamic-Adar index}~\cite{adamic2003friends}. The index was proposed as an alternative to Jaccard for the problem of predicting links, such as friendship or co-authorship, in social networks. 

Link prediction is a widely studied problem, with applications in several Web-related tasks, including hyperlink-prediction~\cite{zhu2002using}, recommender systems in e-commerce~\cite{li2009recommendation}, entity resolution~\cite{malin2005network}, and friend recommendation in social networks~\cite{daud2020applications,freschi2009graph}, among others~\cite{hasan2011survey}. In this problem, each node is characterized by its set of adjacent nodes, or \textit{neighbors}. The intuition is that nodes of similar neighborhoods tend to become neighbors. The Adamic-Adar index is used to compare these sets of neighbors, but unlike Jaccard, it assigns different weights to neighbors (see Section~\ref{sec:background-adamic-adar}). Adamic-Adar is known to be superior to Jaccard for modeling the phenomenon of link emergence in networks in various real-world applications~\cite{liben2007link,martinez2016survey}. The success of Adamic-Adar also motivated the emergence of several metrics with different ways of weighting neighbors~\cite{dong2011link,liu2017extended,cannistraci2013link}, which will be discussed in Section~\ref{sec:link-pred-metrics}.

A second prime example of an application marked by demanding an enormous number of set comparisons is the removal of (near-)duplicate pages in Web search. Eliminating such pages saves network bandwidth, reduces storage costs and improves the quality of search results~\cite{manku2007detecting}. In this domain, each document is commonly treated as a set of word sequences. To find duplicates in a corpus of 10 million pages, a relatively small scale for Web applications~\cite{manku2007detecting}, it would already be necessary to compute the set similarity metric about 50 trillion times. It was precisely in the face of this challenge that the problem of \textit{estimating} set similarity metrics has become highly relevant and has triggered numerous scientific endeavors.


\textit{MinHash}~\cite{broder1997resemblance} and \textit{SimHash}~\cite{charikar2002similarity}, the two best-known estimators, were initially developed for the above-mentioned problem and used respectively in the AltaVista and Google Web search engines. Other current applications taking advantage of set similarity estimation include genomic and metagenomic analysis~\cite{ondov2016mash,koslicki2019improving}, graph comparison~\cite{teixeira2012min}, collaborative filtering~\cite{das2007google}, natural language dataset preparation~\cite{lee2021deduplicating}, and duplicate detection of other types of data such as images~\cite{chum2008near}. 
SimHash is also used in \textit{locality sensitive hashing} (LSH), a technique applied to detect if the similarity of sets exceeds a given threshold, used in problems such as dimensionality reduction, nearest neighbor search, entity resolution and fingerprint matching~\cite{leskovec2020mining}. An important remark is that, despite being related, LSH and the problem addressed in this paper of estimating metrics directly are distinct and both individually relevant as we will discuss in Section~\ref{sec:related}. This wide range of relevant applications illustrates the importance and potential of set similarity estimators to push boundaries of problems where they are applied.


Despite the importance of the estimators mentioned above, previous works reveal limitations of these techniques. As is common with estimators, their accuracy is a function of the input as well as the value to be estimated. In the original MinHash paper, \citet{broder1997resemblance} indicates that the estimator's accuracy is at its worst when the Jaccard index is around 0.5. \citet{koslicki2019improving} show that the probability of the estimator deviating from the true value increases exponentially with the difference in size between the sets to be compared. Nonetheless, \citet{shrivastava2014defense} provides theoretical arguments for the superiority of MinHash over the other baseline, SimHash. More importantly, these techniques are not able to estimate indices like Adamic-Adar. In Section~\ref{sec:experiments}, we will show experimentally that this limitation makes them less appropriate for the link prediction and near-duplicate detection problems. 

Commendable effort has been devoted to reshaping these techniques to overcome these limitations for specific applications as we discuss in Section~\ref{sec:other}. Nevertheless, given its relevance, we argue that it is also important to pursue alternative techniques that explore new perspectives on the problem. With this in mind, we propose \ourhash{}---a novel set similarity estimator. Like its predecessors, \ourhash{} is based on creating a fixed-size representation of sets (often called a \textit{sketch}, \textit{signature} or \textit{fingerprint} in the literature) that can be compared to estimate the similarity between the sets. Creating these compressed representations introduces preprocessing time, but dramatically mitigates the time for each comparison. The central idea of \ourhash{} is to exploit valuable features of high-dimensional random vector spaces, in particular their \textit{tendency towards orthogonality}~\cite{kanerva1988sparse}, to create fixed-dimension sketches while retaining as much information from the original space as possible. In Theorem~\ref{theorem:estimate-intersection-by-dot}, we show that the cardinality of the intersection of two sets can be estimated, without bias, by a simple \textit{dot product} between their \ourhash{} sketches. As the dot product is such a fundamental operation, we argue that \ourhash{} can take advantage of recent progress in modern hardware platforms for enhanced performance and energy efficiency~\cite{yamanaka2008parallel,moughan2021parallel}. 

In addition to the theoretical contribution of a new baseline framework to the problem of comparing sets in its general formulation, we also show that \ourhash{} has prompt practical relevance. We conducted experiments with several popular datasets for the problems of link prediction and near-duplicate detection, which show that the accuracy obtained by \ourhash{} is higher than that obtained with SimHash and MinHash. For this, we exploit the fact that \ourhash{} is able to estimate a more general family of metrics, which are better adapted to applications, as is the case of Adamic-Adar for link prediction. It is worth noting that the time complexity for each comparison is linear in the size of the \ourhash{} sketches regardless of the metric, since it consists of computing the dot product between them. As previously mentioned, these pair-wise comparisons between sketches dictate the overall time complexity.



\section{Background}
\label{sec:background}

In this section, we introduce the most relevant set similarity metrics for the purpose of our work. We start with the Jaccard and Adamic-Adar indices, for which we show theoretical and empirical results. Then, we provide an overview of other metrics that \ourhash{} is able to estimate directly. It is worth noting that many alternative set similarity metrics have been proposed, specifically tailored to better suit specific scenarios by adapting the basic metrics. For an extensive overview of set similarity metrics we recommend \citet{martinez2016survey} and \citet{lu2011link}.

\subsection{Jaccard}
\label{sec:background-jaccard}

Consider the sets $\set{A}$ and $\set{B}$, and let $\set{A} \cap \set{B}$ denote their intersection and $\set{A} \cup \set{B}$ their union. The Jaccard index \cite{jaccard1912distribution} between $\set{A}$ and $\set{B}$ is defined as: 
\begin{align*}
    \jacc{\set{A}, \set{B}}=\frac{\setsize{\set{A} \cap \set{B}}}{\setsize{\set{A} \cup \set{B}}}
\end{align*}
where the vertical bars denote the cardinality of the enclosed set. This is one of the oldest and most established ways of comparing sets. Over time, numerous adaptations of this simple metric have emerged, specializing it for particularly interesting applications. Next, we describe one of these adaptations which is widely used, especially in network science~\cite{barabasi2013network}.


\subsection{Adamic-Adar}
\label{sec:background-adamic-adar}

The Adamic-Adar index was created for the problem of link prediction in social graphs~\cite{adamic2003friends}. Let $G=(\set{V},\set{E})$ be a graph, composed of a set of nodes $\set{V}$ and a set of edges $\set{E}$. Each $e=(u,v)\in \set{E}$ represents an edge (or link) between nodes $u,v\in \set{V}$. With $\neighbors{v} \subseteq \set{V}$ we denote the subset of nodes that are adjacent to $v$, i.e., the \textit{neighbors} of $v$. The cardinality of a neighborhood $|\neighbors{v}|$ is referred to as the \textit{degree} of $v$.

In the context of graphs, Jaccard can be used to compare pairs of nodes by looking at how many connections they have in common, normalized by the size of the union of their neighborhoods. Adamic-Adar seeks to improve this comparison based on the intuition that the more popular a node in the intersection is (i.e., the higher its degree), the less informative it is about the similarity of the nodes being compared. In the case of social networks, for example, a common connection with a celebrity says little about the chance of two people connecting with each other compared to a less popular mutual friend. To account for this, Adamic-Adar penalizes the number of connections that each shared connection has by taking the logarithm of its degree. Formally, the Adamic-Adar between two nodes is defined as:
\begin{align*}
    \adad{u, v} = \sum_{x \in \neighbors{u} \cap \neighbors{v}}\frac{1}{\log{\setsize{\neighbors{x}}}}
\end{align*}

\subsection{Link prediction metrics}
\label{sec:link-pred-metrics}

Given the importance of link prediction, several other metrics have emerged for the purpose of comparing neighborhoods. Similar to Adamic-Adar, these metrics take the local properties of the nodes in the intersection into account. Many of these metrics can also be directly estimated using \ourhash{}. One such example is the Resource Allocation index, used to evaluate the resource transmission between two unconnected nodes through their neighbors which is defined as:
\begin{align*}
\mathrm{RA}({u, v}) = \sum_{x \in \neighbors{u} \cap \neighbors{v}}\frac{1}{\setsize{\neighbors{x}}}
\end{align*}  
In Section~\ref{sec:family-metrics}, we provide a formal description of the family of set similarity metrics that \ourhash{} can directly estimate.

\section{Related Work}
\label{sec:related}



In this section we describe the two most popular estimators, MinHash and SimHash, which will then be used as baselines for the evaluation of \ourhash{}. It is important to highlight that we compare \ourhash{} with these two methods because they constitute the state of the art for the problem in its most general formulation and are used in current applications as already mentioned.


\subsection{MinHash}
\label{sec:minhash}
MinHash~\cite{broder1997resemblance} is a probabilistic method for estimating the Jaccard index. The technique is based on a simple intuition: if we \textit{uniformly sample} one element $x$ from the set $\set{A}\cup \set{B}$, we have that:
\begin{align*}
    \prob{x \in \set{A} \cap \set{B}}=\frac{\setsize{\set{A} \cap \set{B}}}{\setsize{\set{A} \cup \set{B}}}=\jacc{\set{A}, \set{B}}
\end{align*}
which makes the result of this experiment an estimator for Jaccard. However, an important problem remains: \textit{how to uniformly sample from $\set{A}\cup \set{B}$?} Explicitly computing the union is at least as expensive as computing the intersection, that is, it would be as expensive as calculating the Jaccard index exactly. The main merit of MinHash is an efficient way of circumventing this problem.

Let $h: \set{A} \cup \set{B} \to \mathbb{N}$ denote a min-wise independent hash function, i.e., for any subset of the domain, the output of any of its elements is equally likely to be the minimum (see \citet{broder2000min} for a detailed discussion). Then, we have:
\begin{align*}
    \prob{\underset{a \in \set{A}}{\min}\;\hashint{a}=\underset{b \in \set{B}}{\min}\;\hashint{b}} = \frac{\setsize{\set{A} \cap \set{B}}}{\setsize{\set{A} \cup \set{B}}}=\jacc{\set{A}, \set{B}}
\end{align*}
Given the above, the problem of uniformly sampling an element of the union and checking if it belongs to the intersection can be emulated as follows: hash the set elements and check if the smallest value obtained in both sets is the same. Although the result of this random variable is an unbiased estimator of the Jaccard index, its variance is high when the Jaccard is around 0.5. The idea of MinHash is therefore to do $k$ such experiments with independent hash functions and return the sample mean to reduce the variance. 


\subsection{SimHash}
\label{sec:simhash}

SimHash~\cite{charikar2002similarity}, sometimes indistinctly called \textit{angular LSH}, is another popular estimator of set similarity. The sketches of sets are fixed-length binary vectors and are generated as follows: 1) all elements of the superset $S$ are mapped uniformly to vectors in $\{-1,1\}^d$; 2) for each set $X\subseteq S$ a $d$-dimensional vector is created by adding the vectors of its elements; 3) the \textit{SimHash sketch} of the set is a bit string obtained by transforming each positive entry to one and the non-positive entries to zero. The similarity between pairs of sets is then measured by the Hamming distance between these sketches.

Despite being a general estimator for set similarity metrics, SimHash owes its popularity largely to a specific use. \citet{manku2007detecting} showed an efficient way to solve the following problem: \textit{in a collection of SimHash sketches, quickly find all sketches that differ at most $k$ bits from a given sketch, where $k$ is a small integer}. This particular formulation is very useful in the context of duplicate text detection and its efficient solution led to SimHash being used by Google Crawler.

The problem described above is an instance of the problem known as \textit{locality sensitive hashing} (LSH) which, in general, tries to detect pairs of objects whose similarity exceeds a threshold by maximizing the collision probability of the hashes of these objects~\cite{indyk1998approximate,indyk1997locality}. Note that LSH is used to group similar objects and the output is binary, i.e. \textit{two objects are either similar or not}~\cite{leskovec2020mining}. Therefore, we emphasize that the problem of estimating the metrics directly, addressed in this paper, is \textit{different} from LSH. Directly estimating similarity has other possible outcomes, such as ordering pairs by similarity, which is crucial in some applications like query optimization~\cite{cormode2011synopses} and link prediction as we will discuss in Section~\ref{sec:experiments}.

Although SimHash is much more popular in the context of LSH, for the sake of completeness, but underscoring the above, we consider SimHash as a baseline of the general problem as the method was originally proposed by \citet{charikar2002similarity}. Despite this, we make it clear that the other baseline, MinHash, is much more common in the literature for the problem of estimating the actual value of metrics as we discuss in the next section.

\subsection{Adjacent research and developments}
\label{sec:other}

Our primary focus in this study is to address the task of estimating set similarity metrics in its broadest sense. However, it is important to acknowledge other advancements in the field that are not directly aligned with our specific contribution. These advancements, discussed below, primarily involve enhancements tailored to specific contexts and applications.  It is important to note that our method is not intended to directly compete with these notable developments in each individual application, but rather aims to serve as a new baseline for the general problem. We emphasize, however, that \ourhash{} also brings practical contributions to the state of the art. This is achieved by supporting a wider range of metrics, which is formally defined in Section~\ref{sec:family-metrics}. As we will show in Section~\ref{sec:experiments}, this allows for greater accuracy in important problems such as link prediction and document deduplication.

While MinHash remains the standard framework for estimating the actual value of set similarity metrics, several techniques have been proposed to enhance its accuracy and efficiency in specific application contexts. For instance, \citet{chum2012fast} propose an efficient method to compute MinHash sketches for image collections using inverted indexing. Another technique, introduced by \citet{koslicki2019improving}, employs Bloom filters for fast membership queries and is known as \textit{containment MinHash}. They demonstrate the superiority of this technique in metagenomic analysis by more accurately estimating the Jaccard index when dealing with sets that significantly differ in size.

Several other works have focused on a variant of the problem that involves estimating the \textit{weighted} Jaccard index~\cite{wu2020review}. For datasets with sparse data, \citet{ertl2018bagminhash} and \citet{christiani2020dartminhash} have explored the concept of \textit{consistent weighted sampling} (CWS) \cite{ioffe2010improved} with their respective \textit{BagMinHash} and \textit{DartMinHash} techniques. Conversely, when dealing with dense data, methods based on \textit{rejection sampling} have been demonstrated to be more efficient~\cite{shrivastava2016simple,li2021rejection}.

Another important recent endeavor has been to develop LSH techniques based on deep learning, known as ``learn to hash'' methods. These include \textit{Deep Hashing}~\cite{zhu2016deep} and various others~\cite{dutt2019selectivity,hasan2020deep,kipf2018learned}. In general, these techniques do not estimate any particular metric directly, but seek to create sketches that allow for approximate nearest neighbor search. Another key distinction lies in the methodology employed by these approaches. They rely on constructing trained models through annotated data, where the concept of similarity is derived from the mapping of training examples to a specific target. Consequently, this similarity may not necessarily extend to other datasets, making it potentially non-generalizable. In contrast, the methods discussed in this paper estimate the similarity between two sets solely based on the sets themselves. 

Although ``learn to hash'' methods have demonstrated promising accuracy in situations where supervised learning is viable, their broader adoption has also been hindered by other challenges. These limitations encompass high costs associated with training and inference, the inherent unpredictability due to unknown bounds in estimation error, and their high sensitivity to data distribution, often concealed by the reliance on purely empirical assessments~\cite{kim2022learned,wang2021we}. As a result, traditional methods such as MinHash and SimHash continue to be utilized in important applications such as the ones mentioned earlier. Despite the inherent differences and the difficulty of setting up an accurate and unbiased study that delves deep into both approaches, we believe that a comparative study between these traditional methods and learning-based approaches would yield significant value for the scientific community.
\section{DotHash}
\label{sec:ourhash}
We begin by describing a simple method to compute the cardinality of the intersection between two sets. This provides the basis from which we describe the intuition for \ourhash{}. The intuition builds on a generalization of the simple method and a subtle feature of high-dimensional vector spaces. From it, we show how we can create an estimator for the intersection cardinality.
We emphasize that virtually all set similarity metrics are direct functions of the intersection cardinality, combined with other easily obtained quantities such as the size of the sets~\cite{martinez2016survey}. The fact that \ourhash{} estimates the intersection cardinality directly makes it naturally extendable to all these metrics. One of the few exceptions is a family of metrics that assign different weights to the intersection items, such as the Adamic-Adar index. We conclude this section showing how \ourhash{} can be adapted to estimate this larger family of metrics as well, being the first estimator to enable this.



\subsection{Computing the intersection size}
\label{sec:simple_method}

A common way of representing sets is by using bit strings~\cite{rosen2012discrete, fletcher2018comparing}. In this representation, an arbitrary order $[x_1,x_2,\dots,x_d]$ of the elements of the superset $\set{S}$ is established. Then, each set $X \subseteq \set{S}$ is represented by an $\setsize{\set{S}}$-dimensional binary vector whose $i$-th bit is one if $x_i \in X$, and zero otherwise. Table~\ref{tab:bit_strings} illustrates this representation for sets $\set{A} = \{a,b,c,d\}$ and $\set{B} = \{b,c,d,e\}$. This representation is especially common for graphs, where it is called an \textit{adjacency matrix}, and each set consists of the neighborhood of a node. 
\begin{table}[h]
\centering
\caption{Bit string representation of sets $\set{A}$ and $\set{B}$}
\label{tab:bit_strings}
\begin{tabular}{c|ccccc}
        & $a$ & $b$ & $c$ & $d$ & $e$\\
        \midrule
        $\set{A}$ & 1&1&1&1&0\\
        $\set{B}$ & 0&1&1&1&1\\
\end{tabular} 
\end{table}

It is easy to observe that the size of the intersection between $\set{A}$ and $\set{B}$ is given by the number of columns where both elements are one. This provides a straightforward way to get the size of the intersection of sets: calculate the \textit{dot product} between their bit strings \cite{fletcher2018comparing, brusco2021comparison}.




\subsection{Generalization to orthogonal vectors}

\begin{figure}[h]
    \centering
    \resizebox{17em}{!}{
\begin{tikzpicture}[every node/.style={minimum size=1cm-\pgflinewidth, outer sep=2pt}]

\node [draw=none,anchor=south] at (3.5,1.3)
{\fontsize{30}{30}\selectfont $A=\{a,b,c,d\}$};
\foreach \y in {0,...,3}
{
    \node [draw=none,anchor=south] at (-.5,-\y)
    {\fontsize{30}{30}\selectfont$\makealph{\y+1}$};
    \node[fill=black!60] at (.5+\y,.5-\y) {};
    \draw[ultra thick,step=1cm] (0,-\y) grid (5,1-\y);
    \node [draw=none,anchor=south] at (5.5,-\y)
    {\fontsize{25}{25}\selectfont \textbf{...}};
    \draw[ultra thick,step=1cm] (6,0-\y) grid (7,1-\y);
}    

\draw[->,line width=2.5pt] (3.5,-3) -- (3.5,-4.9) node[midway,left] {\fontsize{30}{30}\selectfont \textbf{+}};
\node [draw=none,anchor=south] at (-.5,-6)
{\fontsize{30}{30}\selectfont$A$};
\draw[fill=black!60] (0,-5) rectangle (4,-6);
\draw[ultra thick,step=1cm] (0,-6) grid (5,1-6);
\node [draw=none,anchor=south] at (5.5,-6)
{\fontsize{25}{25}\selectfont \textbf{...}};
\draw[ultra thick,step=1cm] (6,0-6) grid (7,1-6);

\node [draw=none,anchor=south] at (13.5,1.3)
{\fontsize{30}{30}\selectfont $B=\{b,c,d,e\}$};
\foreach \y in {0,...,3}
{
    \node [draw=none,anchor=south] at (9.5,-\y)
    {\fontsize{30}{30}\selectfont$\makealph{\y+2}$};
    \node[fill=black!60] at (11.5+\y,.5-\y) {};
    \draw[ultra thick,step=1cm] (10,-\y) grid (15,1-\y);
    \node [draw=none,anchor=south] at (15.5,-\y)
    {\fontsize{25}{25}\selectfont \textbf{...}};
    \draw[ultra thick,step=1cm] (16,0-\y) grid (17,1-\y);
}    

\draw[->,line width=2.5pt] (13.5,-3) -- (13.5,-4.9) node[midway,left] {\fontsize{30}{30}\selectfont \textbf{+}};
\node [draw=none,anchor=south] at (9.5,-6)
{\fontsize{30}{30}\selectfont$B$};
\draw[fill=black!60] (11,-5) rectangle (15,-6);
\draw[ultra thick,step=1cm] (10,-6) grid (15,1-6);
\node [draw=none,anchor=south] at (15.5,-6)
{\fontsize{25}{25}\selectfont \textbf{...}};
\draw[ultra thick,step=1cm] (16,0-6) grid (17,1-6);

\draw[-,line width=2.5pt] (3.5,-6) -- (3.5,-7);
\draw[-,line width=2.5pt] (13.5,-6) -- (13.5,-7);
\draw[-,line width=2.5pt] (3.5,-7) -- (13.5,-7);
\draw[->,line width=2.5pt] (8.5,-7) -- (8.5,-7.9) node[above=1cm] {\fontsize{40}{40}\selectfont \textbf{.}};
\draw[fill=black!60] (6,-8) rectangle (9,-9);
\draw[ultra thick,step=1cm] (5,-8) grid (10,-9);
\node [draw=none,anchor=south] at (10.5,-9)
{\fontsize{25}{25}\selectfont \textbf{...}};
\draw[ultra thick,step=1cm] (11,-8) grid (12,-9);
\node [draw=none,anchor=south] at (8.5,-10.5)
{\fontsize{30}{30}\selectfont $|A\cap B|=3$};

\end{tikzpicture}
}
    \caption{Intersection calculation using one-hot encoding.}
    \label{fig:dothash-one-hot}
\end{figure}
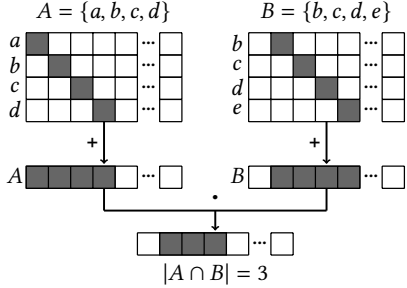

An alternative way of visualizing the set bit strings, important for the generalization we propose, is: consider that each \textit{element} $x_i \in \set{S}$ is encoded by an $\setsize{\set{S}}$-dimensional vector of value one in position $i$, and zero elsewhere. This representation is usually called \textit{one-hot encoding}. From this, we can define the bit string of a set as \textit{the sum of its one-hot encoded elements}, as illustrated in Figure~\ref{fig:dothash-one-hot}.


In Theorem~\ref{theorem:exact-intersection-by-dot}, we show that the dot product results in the intersection of sets not only when they are the sum of one-hot encoded elements, but more generally when we encode the elements using any orthonormal basis of $\mathbb{R}^{\setsize{\set{S}}}$, of which one-hot is a particular case---the \textit{standard basis}. Although this, arguably trivial, generalization alone may not seem advantageous at this point, in the next section we show how it is fundamental in the transition from exact to estimation with our method.



\begin{theorem}
\label{theorem:exact-intersection-by-dot}
Consider arbitrary sets $\set{A},\set{B} \subseteq \set{S}$. Let $\phi: \set{S} \to \mathbb{R}^{\setsize{\set{S}}}$ be any injective map of their elements to vectors to one of the orthonormal bases of $\mathbb{R}^{\setsize{\set{S}}}$, and let
\begin{align*}
    \bm{a} = \sum_{a \in \set{A}} \hashvec{a}\quad\text{and}\quad \; \bm{b} = \sum_{b \in \set{B}} \hashvec{b}\text{.}
\end{align*}
Then, $\bm{a} \cdot \bm{b} = \setsize{\set{A} \cap \set{B}}$.
\end{theorem}

While the above yields a way of representing sets so that we can compute intersection sizes by simple dot products, notice that each set is represented using $\setsize{\set{S}}$ bits, resulting in a time and space complexity of $\bigo{\setsize{\set{S}}}$. Although this can be useful in certain scenarios, clearly this method becomes prohibitively expensive for very large supersets $\set{S}$, for example, in the large scale applications described in the previous sections. Another problem is that in many real applications, such as those related to social networks, the sets change over time, so there is no way to establish the superset size a priori.

In some cases the time and space complexity can be improved to $\bigo{\setsize{\set{A}} + \setsize{\set{B}}}$ by restricting $\hashvec{\cdot}$ to the standard basis encoding, represented using a sparse vector format so that only the non-zero elements are stored. With this modification the dot product can be computed by iterating over both vectors at once, similar to merging lists. However, because of the overhead of sparse vector representations, this is mainly useful when $\setsize{\set{A}} + \setsize{\set{B}} \ll \setsize{\set{S}}$.

There are still limitations to the sparse vector improvement, especially when even $\setsize{\set{A}} + \setsize{\set{B}}$ is very large. This happens both in social networks where nodes can have more than a million connections, and in document deduplication where large documents can hold many word sequences. In the next section we present a method to improve the time complexity to a constant value by giving up the exact intersection size in favour of an estimate.


\subsection{Exploiting \textit{quasi}-orthogonality}
The method described above seems unsuitable for large scale applications as it requires a number of dimensions equal to the size of the superset. This constraint is imposed by the fact that the smallest real vector space with $\setsize{\set{S}}$ orthogonal vectors is $\mathbb{R}^{\setsize{\set{S}}}$. Intuitively, we need orthogonality so that $\hashvec{x}\cdot\bm{a}=1$ if, and only if, $x\in \set{A}$, and zero otherwise. This ensures that $\bm{a} \cdot \bm{b} = \setsize{\set{A} \cap \set{B}}$ (for a detailed discussion see the proof of Theorem~\ref{theorem:exact-intersection-by-dot} in the Appendix~\ref{app:dot-as-intersection-proof}). From an information theory perspective, this guarantees a lossless representation of the sets by the sum of the encoded elements, since by the above operation we can verify exactly which elements make up the set.


Our proposed estimator, \ourhash{}, relies on a very interesting property of high-dimensional vector spaces: \textit{uniformly sampled vectors are nearly orthogonal}, or \textit{quasi}-orthogonal, \textit{to each other with high probability}~\cite{kanerva2009hyperdimensional}. This valuable feature has been explored in several other domains, especially to model human cognition and memory~\cite{kanerva1988sparse,gayler2004vector}.
It is based on this insight that \ourhash{} turns the above method into an estimator for the size of the intersection of sets.

\begin{figure}[h]
    \centering
    \resizebox{18em}{!}{
\begin{tikzpicture}[every node/.style={minimum size=1cm-\pgflinewidth, outer sep=2pt}]

\node [draw=none,anchor=south] at (3.5,1.3)
{\fontsize{30}{30}\selectfont $A=\{a,b,c,d\}$};

\node[fill=c0] at (.5,.5) {};
\node[fill=c4] at (.5+1,.5) {};
\node[fill=c4] at (.5+2,.5) {};
\node[fill=c0] at (.5+3,.5) {};
\node[fill=c4] at (.5+4,.5) {};
\node[fill=c4] at (.5+5,.5) {};
\node[fill=c0] at (.5+6,.5) {};

\node[fill=c0] at (.5,.5-1) {};
\node[fill=c4] at (.5+1,.5-1) {};
\node[fill=c0] at (.5+2,.5-1) {};
\node[fill=c4] at (.5+3,.5-1) {};
\node[fill=c4] at (.5+4,.5-1) {};
\node[fill=c0] at (.5+5,.5-1) {};
\node[fill=c0] at (.5+6,.5-1) {};

\node[fill=c4] at (.5,.5-2) {};
\node[fill=c4] at (.5+1,.5-2) {};
\node[fill=c0] at (.5+2,.5-2) {};
\node[fill=c4] at (.5+3,.5-2) {};
\node[fill=c0] at (.5+4,.5-2) {};
\node[fill=c0] at (.5+5,.5-2) {};
\node[fill=c0] at (.5+6,.5-2) {};

\node[fill=c0] at (.5,.5-3) {};
\node[fill=c4] at (.5+1,.5-3) {};
\node[fill=c4] at (.5+2,.5-3) {};
\node[fill=c0] at (.5+3,.5-3) {};
\node[fill=c4] at (.5+4,.5-3) {};
\node[fill=c0] at (.5+5,.5-3) {};
\node[fill=c0] at (.5+6,.5-3) {};

\foreach \y in {0,...,3}
{
    \node [draw=none,anchor=south] at (-.5,-\y)
    {\fontsize{30}{30}\selectfont$\makealph{\y+1}$};
    \draw[ultra thick,step=1cm] (0,-\y) grid (7,1-\y);
}

\node[fill=c1] at (.5,.5-6) {};
\node[fill=c4] at (.5+1,.5-6) {};
\node[fill=c2] at (.5+2,.5-6) {};
\node[fill=c2] at (.5+3,.5-6) {};
\node[fill=c3] at (.5+4,.5-6) {};
\node[fill=c2] at (.5+5,.5-6) {};
\node[fill=c0] at (.5+6,.5-6) {};

\draw[->,line width=2.5pt] (3.5,-3) -- (3.5,-4.9) node[midway,left] {\fontsize{30}{30}\selectfont \textbf{+}};
\node [draw=none,anchor=south] at (-.5,-6)
{\fontsize{30}{30}\selectfont$A$};
\draw[ultra thick,step=1cm] (0,-6) grid (7,1-6);

\node [draw=none,anchor=south] at (13.5,1.3)
{\fontsize{30}{30}\selectfont $B=\{b,c,d,e\}$};

\node[fill=c0] at (10.5,.5) {};
\node[fill=c4] at (10.5+1,.5) {};
\node[fill=c0] at (10.5+2,.5) {};
\node[fill=c4] at (10.5+3,.5) {};
\node[fill=c4] at (10.5+4,.5) {};
\node[fill=c0] at (10.5+5,.5) {};
\node[fill=c0] at (10.5+6,.5) {};

\node[fill=c4] at (10.5,.5-1) {};
\node[fill=c4] at (10.5+1,.5-1) {};
\node[fill=c0] at (10.5+2,.5-1) {};
\node[fill=c4] at (10.5+3,.5-1) {};
\node[fill=c0] at (10.5+4,.5-1) {};
\node[fill=c0] at (10.5+5,.5-1) {};
\node[fill=c0] at (10.5+6,.5-1) {};

\node[fill=c0] at (10.5,.5-2) {};
\node[fill=c4] at (10.5+1,.5-2) {};
\node[fill=c4] at (10.5+2,.5-2) {};
\node[fill=c0] at (10.5+3,.5-2) {};
\node[fill=c4] at (10.5+4,.5-2) {};
\node[fill=c0] at (10.5+5,.5-2) {};
\node[fill=c0] at (10.5+6,.5-2) {};

\node[fill=c4] at (10.5,.5-3) {};
\node[fill=c0] at (10.5+1,.5-3) {};
\node[fill=c4] at (10.5+2,.5-3) {};
\node[fill=c0] at (10.5+3,.5-3) {};
\node[fill=c0] at (10.5+4,.5-3) {};
\node[fill=c0] at (10.5+5,.5-3) {};
\node[fill=c4] at (10.5+6,.5-3) {};

\foreach \y in {0,...,3}
{
    \node [draw=none,anchor=south] at (9.5,-\y)
    {\fontsize{30}{30}\selectfont$\makealph{\y+2}$};
    \draw[ultra thick,step=1cm] (10,-\y) grid (17,1-\y);
}    

\node[fill=c2] at (10.5,.5-6) {};
\node[fill=c3] at (10.5+1,.5-6) {};
\node[fill=c2] at (10.5+2,.5-6) {};
\node[fill=c2] at (10.5+3,.5-6) {};
\node[fill=c2] at (10.5+4,.5-6) {};
\node[fill=c0] at (10.5+5,.5-6) {};
\node[fill=c1] at (10.5+6,.5-6) {};

\draw[->,line width=2.5pt] (13.5,-3) -- (13.5,-4.9) node[midway,left] {\fontsize{30}{30}\selectfont \textbf{+}};
\node [draw=none,anchor=south] at (9.5,-6)
{\fontsize{30}{30}\selectfont$B$};
\draw[ultra thick,step=1cm] (10,-6) grid (17,1-6);

\draw[-,line width=2.5pt] (3.5,-6) -- (3.5,-7);
\draw[-,line width=2.5pt] (13.5,-6) -- (13.5,-7);
\draw[-,line width=2.5pt] (3.5,-7) -- (13.5,-7);
\draw[->,line width=2.5pt] (8.5,-7) -- (8.5,-7.9) node[above=1cm] {\fontsize{40}{40}\selectfont \textbf{.}};
\node [draw=none,anchor=south] at (8.5,-9.3)
{\fontsize{30}{30}\selectfont $|A\cap B|\approx3$};

\end{tikzpicture}
}
    \caption{\label{fig:dothash-quasi} Intersection calculation using quasi-orthogonal encoding.} 
\end{figure}
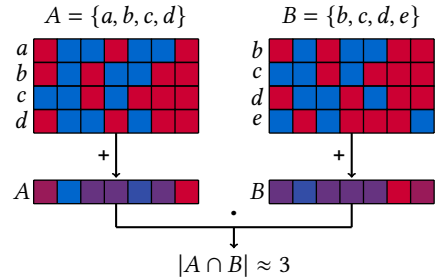

Instead of using a \textit{precisely} orthonormal basis of vectors of $\mathbb{R}^{\setsize{\set{S}}}$ to encode the elements of $\set{S}$, \ourhash{} uses unit vectors sampled from $\mathbb{R}^d$, with $d < \setsize{\set{S}}$. The set \textit{sketches} (fixed-length representations) are then built in the same way, by adding the encodings of their elements, as depicted in Figure~\ref{fig:dothash-quasi}.  Intuitively from the above, each encoded element would be quasi-orthogonal to all others, allowing to approximate the dot product relations mentioned above. In Theorem~\ref{theorem:estimate-intersection-by-dot} we formalize this idea, showing that the the dot product between the \ourhash{} sketches of sets is an unbiased estimator for their intersection cardinality.

\begin{theorem}
\label{theorem:estimate-intersection-by-dot}

Consider arbitrary sets $\set{A},\set{B} \subseteq \set{S}$ and any constant $d \in \mathbb{N}^{+}$. Let $\psi: \set{S} \to \mathbb{R}^d$ be a uniform random mapping of elements in $\set{S}$ to unit vectors which are the vertices of a $d$-dimensional hypercube, and let
\begin{align*}
    \bm{a} = \sum_{a \in \set{A}} \hashqvec{a}\quad\text{and}\quad \; \bm{b} = \sum_{b \in \set{B}} \hashqvec{b}\text{.}
\end{align*}
Then, $\expected{\bm{a} \cdot \bm{b}} = \setsize{\set{A} \cap \set{B}}$, and
\begin{align*}
    \variance{\bm{a} \cdot \bm{b}} = \frac{1}{d}\left(\setsize{\set{A}} \setsize{\set{B}} + \setsize{\set{A} \cap \set{B}}^2 - 2\setsize{\set{A} \cap \set{B}}\right)
\end{align*}
\end{theorem}

\begin{figure}[h]
 \centering
 \scalebox{1.0}{\subimport*{figures/}{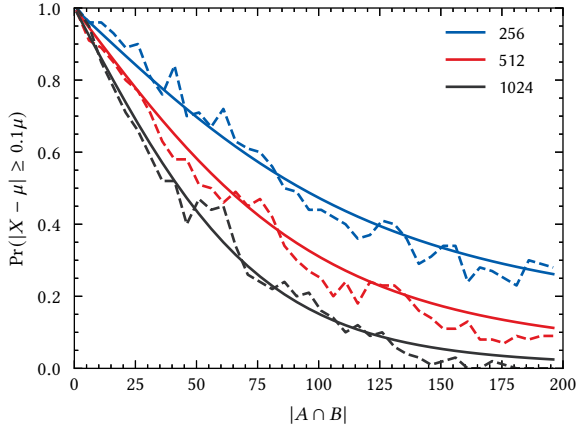}}
  \caption{\label{fig:intersection-estimate-bound} Intersection estimate bounds for \ourhash{}. With $|\set{A}| = |\set{B}| = 200$ and $d=\{256, 512, 1024\}$. Dashed lines indicate experimental results.} 
\end{figure}

Using the variance provided in Theorem~\ref{theorem:estimate-intersection-by-dot} and the Chebychev inequalty we can bound the probablity of error by:
\begin{align*}
    \prob{\abs{X - \mu} \geq \epsilon \mu} \leq \frac{\variance{\bm{a}\cdot\bm{b}}}{\left(\epsilon \setsize{\set{A} \cap \set{B}}\right)^2} 
\end{align*}
If we use the observation that each dimension can be interpreted as an independent sample, we can use the Central Limit Theorem (CLT) to approximate the probability of error as follows:
\begin{align*}
    \lim_{d\to\infty} \prob{\abs{X - \mu} \geq \epsilon \mu} = 2 \left(1 -\normalcdf{\frac{\epsilon \setsize{\set{A} \cap \set{B}}}{\sqrt{\variance{\bm{a}\cdot\bm{b}}}}}\right) 
\end{align*}
where $\normalcdf{\cdot}$ denotes the standard normal cumulative distribution function. 
In Figure~\ref{fig:intersection-estimate-bound} we provide the CLT estimate (solid line) and the empirical probability (dashed line).

We can rewrite the CLT (or the Chebychev inequality) to get the required number of dimensions $d$ to obtain an error greater or equal to $\epsilon \setsize{\set{A} \cap \set{B}}$ with a given probability $p$:
\begin{align*}
 d \approx \variance{\bm{a}\cdot\bm{b}}\left(\frac{\normalppf{1 - \frac{p}{2}}}{\epsilon \setsize{\set{A} \cap \set{B}}} \right)^2
\end{align*}
where $\normalppf{\cdot}$ denotes the standard normal percent point function.

\subsection{Estimating Adamic-Adar}

Now that we have established a method for estimating the size of the intersection, we describe how to adapt \ourhash{} to estimate the Adamic-Adar index. 
The idea starts from the fact that $\bm{a}$ and $\bm{b}$ are sums of the vectors that encode the elements of their respective sets. Given this and the distributive property of the dot product over addition, we have:
\begin{align*}
    \bm{a} \cdot \bm{b} = \sum_{a \in \set{A}} \sum_{b \in \set{B}} \hashqvec{a} \cdot \hashqvec{b}
\end{align*}
Observing that $\expected{\hashqvec{a} \cdot \hashqvec{b}}=1$ if $a=b$, and zero otherwise (see the proof of Theorem~\ref{theorem:estimate-intersection-by-dot}):
\begin{align*}
    \label{eq:def-intersect-size}
    \expected{\bm{a} \cdot \bm{b}}=\sum_{a \in \set{A}} \sum_{b \in \set{B}} \expected{\hashqvec{a} \cdot \hashqvec{b}} = \sum_{x\in \set{A} \cap \set{B}} 1
\end{align*}
The right-hand side of this equation is similar to Adamic-Adar in that both sum values over the intersection items. The key missing part is that the value to be summed must be a function of the size of the neighborhoods, not a constant.

In the above case, the summation parameter is one because every element is encoded to a unit vector. However, the construction of \ourhash{} allows us to adjust the summation parameter by modifying the magnitude of the vectors used to represent each element. To obtain the Adamic-Adar index, we want each element, in this case each \textit{node}, to be encoded in such a way that:
\begin{align*}
    \hashqvec{v} \cdot \hashqvec{v} = \frac{1}{\log{\setsize{\neighbors{v}}}}
\end{align*}
Theorem~\ref{theorem:estimate-adar} shows how to adapt the vector magnitudes to obtain the Adamic-Adar index.

\begin{figure*}[ht]
 \centering
 \scalebox{0.8}{\subimport*{figures/}{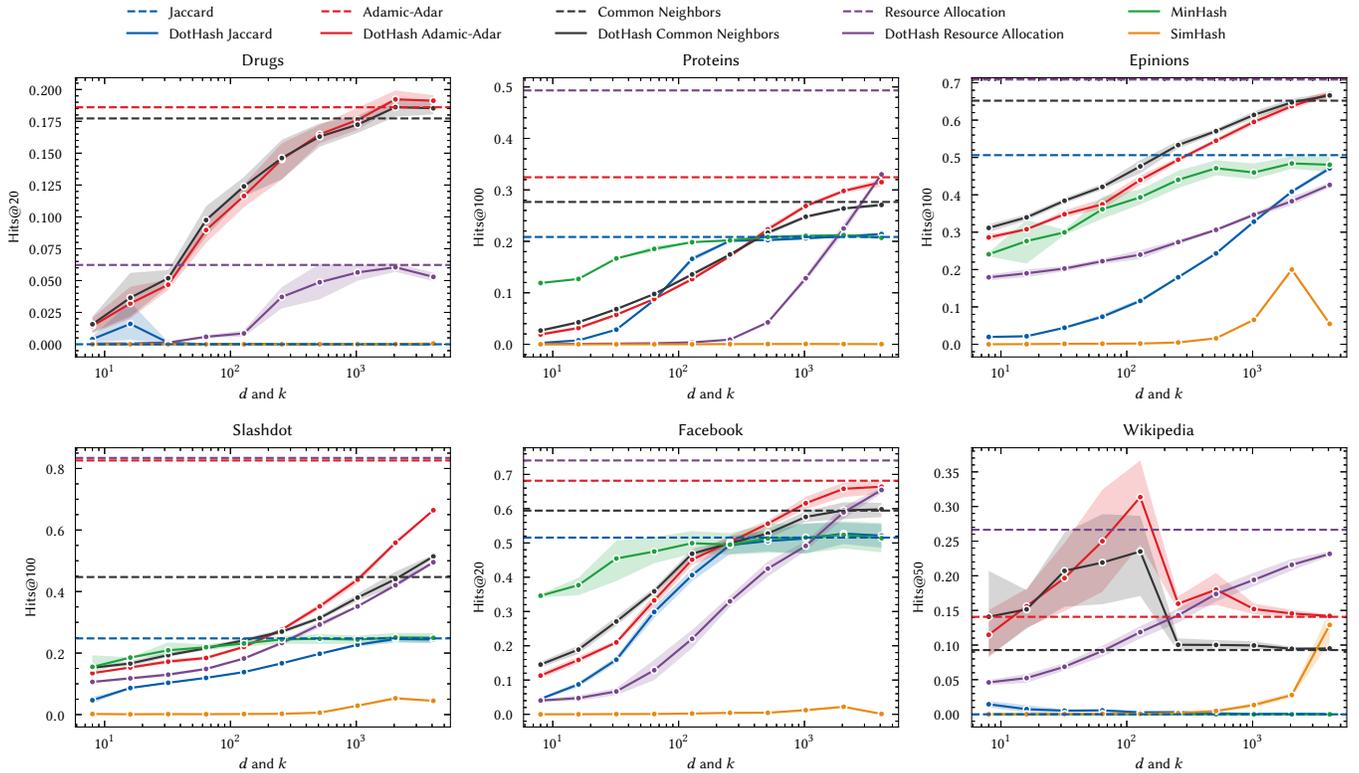}}
  \caption{\label{fig:link-pred-hits} Link prediction accuracy results while varying the number of dimensions $d$ and hashes $k$.} 
\end{figure*}

 \begin{theorem}
 \label{theorem:estimate-adar}

 Consider an arbitrary graph $G = (\set{V}, \set{E})$ with nodes $\set{V}$ and edges $\set{E}$. Take any constant $d \in \mathbb{N}^{+}$ and let $\neighbors{v} \subseteq \set{V}$ denote the neighbors of node $v \in \set{V}$. Let $\psi: \set{V} \to \mathbb{R}^d$ be a uniform random mapping of nodes in $\set{V}$ to unit vectors which are the vertices of a $d$-dimensional hypercube, and let
 \begin{align*}
     \bm{u} =& \sum_{x \in \neighbors{u}} \hashqvec{x}\sqrt{\frac{1}{\log{\setsize{\neighbors{x}}}}}\\
 \end{align*}
 and
\begin{align*}
     \bm{v} =& \sum_{y \in \neighbors{v}} \hashqvec{y}\sqrt{\frac{1}{\log{\setsize{\neighbors{y}}}}}.
\end{align*}
Then, $\expected{\bm{u} \cdot \bm{v}} = \adad{u, v}$.
\end{theorem}


\subsection{General family of supported metrics}
\label{sec:family-metrics}

Building upon the result presented in the previous section, we naturally extend it to encompass a general formulation of all set similarity metrics that can be directly estimated using \ourhash{}. This family includes all metrics of the form: $\sum_{x \in \set{A} \cap \set{B}} f(x)$,
where $f : \set{S} \to \mathbb{R}$ is any function on intersection elements. Besides the item $x$, the function $f$ can use any global parameters such as the cardinalities of $\set{A}$, $\set{B}$ or $\set{S}$. The \ourhash{} sketches for this general set similarly metric are given by:
\begin{align*}
    \bm{a} = \sum_{a \in \set{A}} \hashqvec{a} \sqrt{f(a)} \quad \text{and} \quad \bm{b} = \sum_{b \in \set{B}} \hashqvec{b} \sqrt{f(b)}
\end{align*}
and the estimate is obtained by $\bm{a} \cdot \bm{b}$. Observe that the intersection size, the Adamic-Adar index, and the Resource Allocation index all fit into this general framework: the intersection size corresponds to $f(x) = 1$, Adamic-Adar to $f(x) = \frac{1}{\log{\setsize{\neighbors{x}}}}$, and Resource Allocation to $f(x) = \frac{1}{\setsize{\neighbors{x}}}$. This group of metrics directly supported by \ourhash{} includes the majority of metrics listed in~\citet{martinez2016survey} and ~\citet{lu2011link}.
\section{Experiments}
\label{sec:experiments}

In this section we present experiments comparing \ourhash{} to the baselines presented in Section~\ref{sec:related}. The main goal here is to provide empirical evidence on the advantages of \ourhash{} in the link prediction and duplicate detection tasks.
All the methods were implemented using PyTorch~\cite{paszke2019pytorch} and the Torchhd library \cite{heddes2022torchhd}, and ran on a machine with 20 Intel Xeon Silver 4114 CPUs, 93 GB of RAM and 4 Nvidia TITAN Xp GPUs. The experiments, however, only used a single CPU or GPU. We repeated each experiment 5 times on the CPU and 5 times on the GPU. 
The code is available at: \url{https://github.com/mikeheddes/dothash}.

\begin{table}[h]
    \centering
    \caption{Statistics of the graph datasets}
    \label{table:datasets_stats}
    \begin{tabular}{l|lll}
        \toprule
        Dataset & Nodes & Edges & Median degree\\
        \midrule
        Drugs & 4,267 & 1,334,889 & 446 \\
        Wikipedia & 11,631 & 341,691 & 13 \\
        Facebook & 22,470 & 342,004 & 7 \\
        Epinions & 75,879 & 508,837 & 2\\
        Slashdot & 82,168 & 948,464 & 6\\
        Proteins & 576,289 & 42,463,862 & 43\\
        \bottomrule
    \end{tabular}
\end{table}

\subsection{Datasets}
\label{sec:datasets}

An overview of the datasets used is shown in Table~\ref{table:datasets_stats}. To compare the methods under different circumstances, we consider a range of common benchmarks used in the literature that have different characteristics and are associated with different applications. For the link-prediction task we evaluate each method on the following datasets:

\begin{itemize}
    \item Drugs~\cite{inproceedings}: This dataset represents the interaction between drugs, where the joint effect of using both drugs is significantly different from their effects separately.
    \item Wikipedia~\cite{wiki}: This dataset represents a webpage network where each node represents a web page and the edges represent hyperlinks between them.
    \item Facebook~\cite{wiki}: A network of verified Facebook pages where nodes correspond to Facebook pages and edges are the mutual likes between pages.
    \item Proteins~\cite{Szklarczyk2019STRINGVP}: Is a protein network where nodes represent proteins from different species and edges show biological meaningfulness between the proteins associations.
    \item Epinions~\cite{10.1007/978-3-540-39718-2_23}: Represents the who-trusts-whom social network of the general consumer review site Epinions.com, where each node represents a user, and each edge is a directed trust relation.
    \item Slashdot~\cite{leskovec2008community}: Represents the Slashdot social network as of February 2009, where each node is a user and each edge is a directed friend/foe link.
\end{itemize}

For the document deduplication we use the following datasets:
\begin{itemize}
    \item CORE Deduplication Dataset 2020~\cite{dedup2020}: This dataset consists of more than 1.3M scholar documents labeled as duplicates or non-duplicates. 
    \item Harvard Dataverse Duplicate Detection Restaurant dataset~\cite{DVN/O7LNDH_2022}. This dataset consists of a collection of 864 restaurant records containing 112 duplicates.
    \item Harvard Dataverse Duplicate Detection cddb dataset~\cite{DVN/O7LNDH_2022}: This dataset contains a set of 9,763 records with 299 duplicated entries, with each row representing information about a particular audio recording. 
\end{itemize}

\subsection{Link prediction}


This experiment aims to demonstrate a practical advantage of \ourhash{}. While the baseline estimators, MinHash and SimHash, are limited to estimating the Jaccard index, \ourhash{} offers the ability to estimate more complex metrics, known to be more effective for certain applications, including link prediction. We evaluate the accuracy of each estimator in solving the link prediction problem. The problem consists of inferring which links will appear in the future, called the \textit{inference time interval}, given a snapshot of a graph~\cite{liben2007link}. In practice, the task is seen as a ranking problem for pairs of nodes, i.e., the approaches compare pairs of nodes and predict that the most similar pairs are those that are likely to connect in the future.

The quality of the methods is evaluated based on how well they rank pairs that effectively form in the inference time interval, against random pairs that do not connect. These edges are respectively called the positive and negative samples. The most popular metric, Hits@$K$, counts the ratio of positive edges that are ranked $K$ or above the negative edges \cite{hu2020open}. The Jaccard, Adamic-Adar, Common Neighbors, and Resource Allocation indices are all used in the literature for establishing this ranking. 

\begin{figure}[h]
\centering
\scalebox{0.74}{\subimport*{figures/}{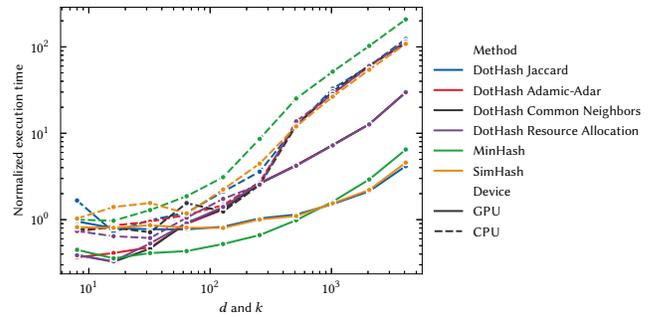}}
 \caption{\label{fig:lp-norm-time} Normalized average execution time of different methods, relative to MinHash with $k=8$ running on CPU. The average is calculated over all link prediction datasets.} 
\label{fig:exec_time_LP}
\end{figure}

The results, presented in Figure~\ref{fig:link-pred-hits}, provide evidence to substantiate the claim that estimating more appropriate metrics makes \ourhash{} a better estimator for the link prediction problem. By employing sufficient dimensions and selecting suitable metrics, \ourhash{} consistently outperforms the baselines across all datasets and approaches the exact indices, shown in dashed lines. Each solid line represents the mean of the values observed in the five repetitions, and the corresponding colored shades show the 95\% confidence interval. Importantly, as explained in the previous section, the adoption of \ourhash{} does not impose a significant additional computational burden compared to the baseline methods. Figure~\ref{fig:exec_time_LP} shows the normalized execution time of each method on the same datasets. For more details on execution times, please refer to Appendix~\ref{app:detailed_exps}.

\subsection{Document deduplication}
\label{sec:exp_deduplication}
The detection of duplicate documents was the first major application to motivate the development of set similarity estimators. Both MinHash and SimHash were developed and became popular for their use in deduplicating web pages in Google and Alta Vista search engines. With our experiments in this section we seek to show how \ourhash{} compares to these methods in this important problem.

Once again, we took advantage of the broader capability of \ourhash{} to estimate a metric richer than the Jaccard coefficient between documents. While MinHash and SimHash give equal weight to shingles (sequences of words) when comparing documents, with \ourhash{} we can assign different weights to reflect how important each shingle is to each document in the corpus. Our intuition is that the more important a term is to identify each text, the more its presence in different texts indicates their similarity. 

One of the most popular ways of evaluating how important a term is to a document is by the \textit{inverse document frequency}, or IDF~\cite{jones1972statistical}. The measure is widely used in the information retrieval literature for text mining~\cite{robertson2004understanding,henzinger2003query}, and is defined as:
\begin{align*}
    \idf{x} = \log{\frac{|D|}{|\{d \in D: x \in d\}|}}
\end{align*}
where $|D|$ is the number of documents in the entire corpus and $|\{d \in D: x \in d\}|$ the number of documents that contain the term $x$. Given this, we can compare documents $A$ and $B$ not only by the number, but also by the importance of common shingles, as follows:
\begin{align*}
    \mathrm{sim_{idf}}(A,B) = \sum_{x\in A\cap B}\idf{x}
\end{align*}
which is in the family of functions that \ourhash{} can estimate.

In Table~\ref{table:deduplication_results} we show the comparative results between the three estimators for the near-duplicate detection problem in the three different datasets described in Section~\ref{sec:datasets}. For each estimator we present the near-duplicity detection accuracy in terms of Hits@25 and execution time in seconds. The number of hash values and dimensions for MinHash, SimHash, and \ourhash{} were set to 128, 500, and 10,000, respectively. These values were chosen to ensure comparable accuracy and enable the observation of differences in computational efficiency between the algorithms.

The numerical results presented indicate that \ourhash{} is able to surpass the accuracy of MinHash in all datasets, even with a number of dimensions in which its execution time is between 0.5 and 3$\times$ faster. The same is observed in the comparison with SimHash, which obtains the lowest accuracy in all cases. These empirical results reinforce the findings observed in the link prediction experiments, highlighting the advantage of \ourhash{}. By efficiently estimating richer metrics through a single dot product computation between set sketches, \ourhash{} consistently delivers superior cost-benefit compared to other estimators.

\begin{table}[h]
    \centering
\caption{Accuracy and computation time (in seconds) results in the detection of duplicate documents.}
\label{table:deduplication_results}
\begin{tabular}{ll|ccc}
        \toprule
Dataset     &   Metric       & MinHash  & SimHash  &  DotHash     \\
        \midrule

CORE 20' & Hits@25 & 0.6246   &  0.3991  &   \textbf{0.6286}    \\
        & Time     & 0.0269  &  0.0284 &  \textbf{0.0088} \\[3pt]
Rest & Hits@25 &   0.9598    &  0.7745    &  \textbf{0.9819}   \\
        & Time & 0.0014  & 0.0011   &  \textbf{0.0006}   \\[3pt]
cddb & Hits@25 & 0.9058    & 0.6496   & \textbf{0.9085}     \\
        & Time    &  0.0094  & 0.0084 & \textbf{0.0066}\\
        \bottomrule
\end{tabular}
\end{table}

\section{Conclusion}
\label{sec:conclusion}

We propose \ourhash{}, a new baseline method for estimating the similarity between sets. The method takes advantage of the tendency to orthogonality of sets of random high-dimensional vectors to create fixed-size representations for sets. We show that a simple dot product of these sketches serves as an unbiased estimator for the size of the intersection of sets. \ourhash{} allows estimating a larger set of metrics than existing estimators. Our experiments show that this makes it more appropriate for link prediction and duplicate detection tasks. Adding the theoretical and practical contribution, we see \ourhash{} as a new framework for a problem of increasing relevance in data mining and related areas.



\bibliographystyle{ACM-Reference-Format}
\bibliography{refs}

\clearpage
\appendix


\section{Theorem Proofs}
\label{app:thm_proofs}
In this section we provide the proofs of the theorems presented in the main paper.

\subsection{Intersection cardinality by dot product}
\label{app:dot-as-intersection-proof}

\begin{customthm}{1}
Consider arbitrary sets $A,B \subseteq S$. Let $\phi: S \to \mathbb{R}^{\setsize{S}}$ be any injective map of their elements to vectors in one of the orthonormal bases of $\mathbb{R}^{\setsize{S}}$, and let
\begin{align*}
    \bm{a} = \sum_{a \in A} \hashvec{a}\quad\text{and}\quad \; \bm{b} = \sum_{b \in B} \hashvec{b}\text{.}
\end{align*}
Then, $\bm{a} \cdot \bm{b} = \setsize{A \cap B}$.
\end{customthm}

\begin{proof}
By the definitions of $\bm{a}$ and $\bm{b}$,
\begin{align*}
    \bm{a} \cdot \bm{b} &= \left( \sum_{a \in A} \hashvec{a} \right) \cdot \left( \sum_{b \in B} \hashvec{b} \right)
\end{align*}
Using the distributivity of the dot product over addition, we can rewrite the above equation as:
\begin{align*}
    \bm{a} \cdot \bm{b} = \sum_{a \in A} \sum_{b \in B} \hashvec{a} \cdot \hashvec{b}\\
\end{align*}
Because $\phi$ is injective, i.e., $\hashvec{a} = \hashvec{b}$ if, and only if, $a=b$, and based on the orthogonal property of orthonormal vectors, $\hashvec{a} \cdot \hashvec{b} = 1$ if $a=b$, and zero otherwise. Then,
\begin{align*}
    \bm{a} \cdot \bm{b} = \sum_{a \in A} \sum_{b \in B} \indicator{a = b} = \sum_{x \in A \cap B} 1 =  \setsize{A \cap B}
\end{align*}
\end{proof}

\subsection{Intersection cardinality estimate by dot product}

\begin{customthm}{2}
Consider arbitrary sets $A,B \subseteq S$ and any constant $d \in \mathbb{N}^{+}$. Let $\psi: S \to \mathbb{R}^d$ be a uniform random mapping of elements in $S$ to unit vectors which are the points of a $d$-dimensional hypercube, and let
\begin{align*}
    \bm{a} = \sum_{a \in A} \hashqvec{a}\quad\text{and}\quad \; \bm{b} = \sum_{b \in B} \hashqvec{b}\text{.}
\end{align*}
Then, $\expected{\bm{a} \cdot \bm{b}} = \setsize{A \cap B}$, and
\begin{align*}
    \variance{\bm{a} \cdot \bm{b}} = \frac{1}{d}\left(\setsize{A} \; \setsize{B} + \setsize{A \cap B}^2 - 2\setsize{A \cap B}\right)
\end{align*}
\end{customthm}

\begin{proof}
By the definitions of $\bm{a}$ and $\bm{b}$,
\begin{align*}
    \bm{a} \cdot \bm{b} = \left( \sum_{a \in A} \hashqvec{a} \right) \cdot \left( \sum_{b \in B} \hashqvec{b} \right)
\end{align*}
Applying the linearity of expectation and distributivity of the dot product over addition, we have:
\begin{align*}
    \expected{\bm{a} \cdot \bm{b}} = \expected{\sum_{a \in A} \sum_{b \in B} \hashqvec{a} \cdot \hashqvec{b}}
    = \sum_{a \in A} \sum_{b \in B} \expected{\hashqvec{a} \cdot \hashqvec{b}}
\end{align*}
Given that $\psi$ maps uniformly to unit vectors which are the points of a $d$-dimensional hypercube, the elements of $\hashqvec{\cdot}$ are sampled uniformly at random from $\left\{-\frac{1}{\sqrt{d}}, +\frac{1}{\sqrt{d}}\right\}$. When $a = b$,
\begin{align*}
    \expected{\hashqvec{a} \cdot \hashqvec{b}} = \expected{\hashqvec{a} \cdot \hashqvec{a}} = \expected{\sum^d_{i=1} \hashqvec{a}_i^2} = \sum^d_{i=1} \frac{1}{d} = 1   
\end{align*}
where the right-hand subscripts denote the dimension of the vector. And when $a \neq b$,
\begin{align*}
    \expected{\hashqvec{a} \cdot \hashqvec{b}} = \expected{\sum^d_{i=1} \hashqvec{a}_i \; \hashqvec{b}_i} = \sum^d_{i=1} \expected{\hashqvec{a}_i \; \hashqvec{b}_i} = 0   
\end{align*}
Thus, $\expected{\hashqvec{a}\cdot\hashqvec{b}}=\indicator{a = b}$, which ensures that:
\begin{align*}
    \expected{\bm{a} \cdot \bm{b}} = 
    \sum_{a \in A} \sum_{b \in B} \indicator{a = b} = \sum_{x \in A \cap B} 1 = \setsize{A \cap B}
\end{align*}
Moreover, the variance of the estimator is obtained as follows, where $r_i\left(a, b\right) = \hashqvec{a}_i \; \hashqvec{b}_i$ and we use the definitions of $\bm{a}$, $\bm{b}$, and the dot product,

\begin{align*}
    \variance{\bm{a} \cdot \bm{b}} = \variance{\sum_{i = 1}^{d} \sum_{a \in A} \sum_{b \in B} r_i\left(a, b\right)}
\end{align*}
Since each dimension of $\hashqvec{\cdot}$ is sampled independently from an identical distribution (i.i.d.),
\begin{align*}
    \variance{\bm{a} \cdot \bm{b}} = \sum_{i = 1}^{d} \variance{\sum_{a \in A} \sum_{b \in B} r_i\left(a, b\right)}
\end{align*}
We then separate the equal from the non-equal pairs of $a$ and $b$ and note that when $a = b$ their elements are identical thus their outcome has no variance, this gives:
\begin{align*}
    \variance{\bm{a} \cdot \bm{b}} =& \sum_{i = 1}^{d} \variance{\sum_{x \in A \cap B} \hashqvec{x}_i^2} + \variance{\sum_{a \in A} \sum_{b \in B \setminus \{a\}} r_i\left(a, b\right)}\\
    =& \sum_{i = 1}^{d} \variance{\sum_{a \in A} \sum_{b \in B \setminus \{a\}} r_i\left(a, b\right)}
\end{align*}
Using the property of linear combination of random variables we get,
\begin{align*}
    \variance{\bm{a} \cdot \bm{b}} &= \sum_{i = 1}^{d} \sum_{a \in A} \sum_{b \in B \setminus \{a\}} \variance{r_i\left(a, b\right)}\\ &+ \sum_{x \in A \setminus \{a\}} \sum_{y \in B \setminus \{x, b\}} \covariance{r_i\left(a, b\right), r_i\left(x, y\right)}
\end{align*}
Lastly, we observe that the covariance is $\frac{1}{d^2}$ when $(a, b) = (y, x)$ giving the variance:
\begin{align*}
    \variance{\bm{a} \cdot \bm{b}} =& \sum_{i = 1}^{d} \frac{1}{d^2}\left(\setsize{A} \; \setsize{B} + \setsize{A \cap B}^2 - 2 \setsize{A \cap B} \right)\\
    =& \frac{1}{d} \left(\setsize{A} \; \setsize{B} + \setsize{A \cap B}^2 - 2 \setsize{A \cap B} \right)
\end{align*}
\end{proof}

\begin{figure*}[t]
 \centering
 \scalebox{0.7}{\subimport*{figures/}{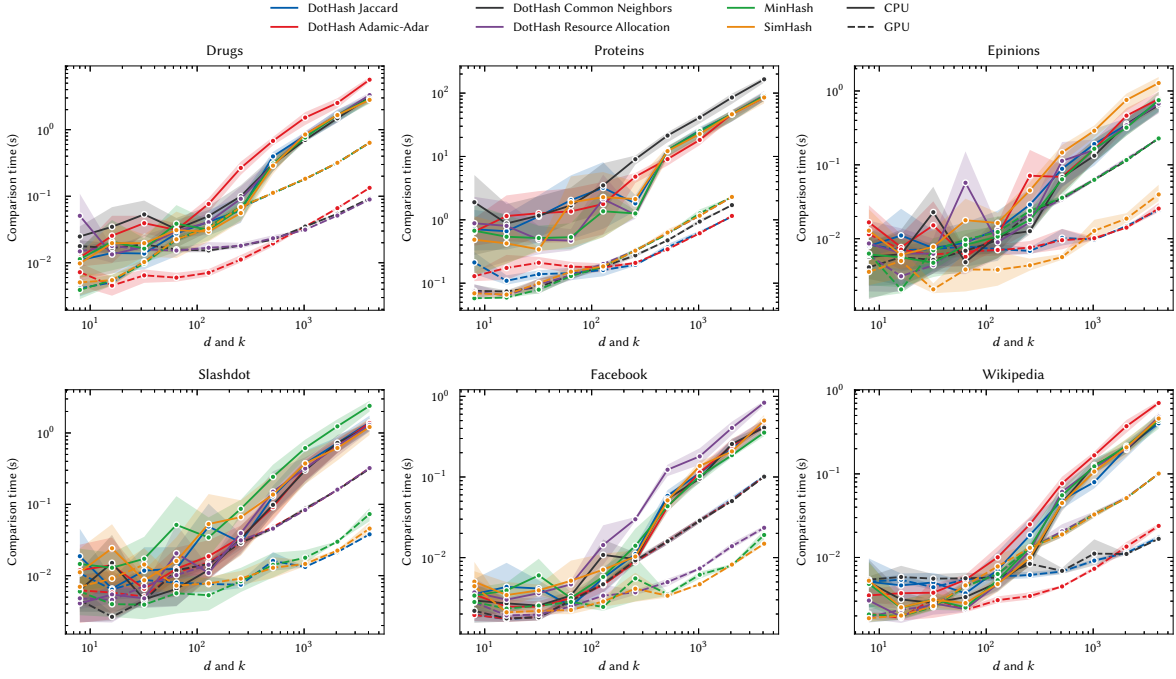}}
  \caption{Node signature comparison time in the link prediction tasks.}
  \label{fig:link-pred-time} 
\end{figure*}

\subsection{Adamic-Adar estimate by dot product}

\begin{customthm}{3}

Consider an arbitrary graph $G = (V, E)$ with nodes $V$ and edges $E$. Take any constant $d \in \mathbb{N}^{+}$ and let $\neighbors{v} \subseteq V$ denote the neighbors of node $v \in V$. Let $\psi: V \to \mathbb{R}^d$ be a uniform random mapping of nodes in $V$ to unit vectors pointing to the corners of a $d$-dimensional hypercube, and let
\begin{align*}
 \bm{u} = \sum_{x \in \neighbors{u}} \hashqvec{x}\sqrt{\frac{1}{\log{\setsize{\neighbors{x}}}}}\quad\text{and}\quad \;
     \bm{v} = \sum_{y \in \neighbors{v}} \hashqvec{y}\sqrt{\frac{1}{\log{\setsize{\neighbors{y}}}}}
\end{align*}
Then, $\expected{\bm{u} \cdot \bm{v}} = \adad{u, v}$.
\end{customthm}
\begin{proof}
By the definitions of $\bm{u}$ and $\bm{v}$,
\begin{align*}
    \bm{u} \cdot \bm{v} = \left(\sum_{x \in \neighbors{u}} \hashqvec{x}\sqrt{\frac{1}{\log{\setsize{\neighbors{x}}}}}\right) \cdot \left(\sum_{y \in \neighbors{v}} \hashqvec{y}\sqrt{\frac{1}{\log{\setsize{\neighbors{y}}}}}\right)
\end{align*}
Applying the linearity of expectation and distributivity of the dot product over addition, we have:
\begin{align*}
    &\expected{\bm{u} \cdot \bm{v}}\\ 
    &= \expected{\sum_{x \in \neighbors{u}} \sum_{y \in \neighbors{v}} \left(\hashqvec{x}\sqrt{\frac{1}{\log{\setsize{\neighbors{x}}}}}\right) \cdot \left(\hashqvec{y}\sqrt{\frac{1}{\log{\setsize{\neighbors{y}}}}}\right)}\\
    &= \sum_{x \in \neighbors{u}} \sum_{y \in \neighbors{v}} \expected{\left(\hashqvec{x}\sqrt{\frac{1}{\log{\setsize{\neighbors{x}}}}}\right) \cdot \left(\hashqvec{y}\sqrt{\frac{1}{\log{\setsize{\neighbors{y}}}}}\right)}\\
    &= \sum_{x \in \neighbors{u}} \sum_{y \in \neighbors{v}}\sqrt{\frac{1}{\log{\setsize{\neighbors{x}}}}}\sqrt{\frac{1}{\log{\setsize{\neighbors{y}}}}} \expected{\hashqvec{x} \cdot \hashqvec{y}}
\end{align*}
From the previous proof we have that $\expected{\hashqvec{x}\cdot\hashqvec{y}}=\indicator{x = y}$, giving:
\begin{align*}
    &\expected{\bm{u} \cdot \bm{v}}\\
    &= \sum_{x \in \neighbors{u}} \sum_{y \in \neighbors{v}}\sqrt{\frac{1}{\log{\setsize{\neighbors{x}}}}}\sqrt{\frac{1}{\log{\setsize{\neighbors{y}}}}}\indicator{x = y}\\
    &= \sum_{x \in \neighbors{u}\cap\neighbors{v}}\frac{1}{\log{\setsize{\neighbors{x}}}}=\adad{u,v}
\end{align*}
\end{proof}

\section{Detailed Experimental Results}
\label{app:detailed_exps}

In this section we present additional experimental results. We believe that the results presented in the main paper are sufficient to substantiate our main claims. Still, the results below should reveal more details to those interested in, for example, exploring the \ourhash{} framework in future work.

\subsection{Time to compare node signatures in link prediction}
The results in the Figure~\ref{fig:link-pred-time} show the time taken to compare the sketch of the sets created by each method in the link prediction datasets. For each of the methods and datasets, we show the CPU and GPU times for different sketch dimensions $d$, in the case of DotHash and SimHash, and minwise hash functions $k$ for MinHash. The lines represent the mean over 5 runs of calculating the represented metric for all edges in the test set.

\end{document}